\documentclass[aps,prd,floats,floatfix, twocolumn,amssymb,amsmath,
superscriptaddress,nofootinbib,showpacs,longbibliography]{revtex4-2}

\usepackage[T1]{fontenc}
\usepackage[utf8]{inputenc}
\usepackage{txfonts} 
\usepackage[normalem]{ulem} 

\usepackage{graphicx}
\usepackage[linktocpage,breaklinks]{hyperref}
\usepackage[capitalize]{cleveref}
\usepackage[usenames,dvipsnames]{xcolor}
\hypersetup{colorlinks=true,citecolor=NavyBlue,
linkcolor=NavyBlue,urlcolor=NavyBlue}

\usepackage{multirow,array}
\DeclareMathAlphabet{\pazocal}{OMS}{zplm}{m}{n}
\usepackage{microtype}
\usepackage{subfigure}
\usepackage{journals}

\newcommand{\dd}{{\rm d}}
\newcommand{\nn}{\nonumber\\}

\newcommand{\hust}{\affiliation{School of Physics, Huazhong University of Science and Technology, Wuhan 430074, China}}

\begin{document}

\title{Difference between quark stars and neutron stars in universal relations and their effect on gravitational waves} 

\author{Duanyuan Gao}
\hust

\author{Hao-Jui Kuan}
\affiliation{Department of Physics, The Grainger College of Engineering, University of Illinois Urbana-Champaign, Urbana, IL 61801, USA}
\affiliation{Max Planck Institute for Gravitational Physics (AEI), Potsdam/Golm, Germany}

\author{Yurui Zhou}
\hust

\author{Zhiqiang Miao}
\affiliation{Tsung-Dao Lee Institute, Shanghai Jiao Tong University, Shanghai, China}

\author{Yong Gao}
\affiliation{Max Planck Institute for Gravitational Physics (AEI), Potsdam/Golm, Germany}

\author{Chen Zhang}
\affiliation{School of Physics Science and Engineering, Tongji University, Shanghai, China
}

\author{Enping Zhou}
\email{ezhou@hust.edu.cn}
\hust

\date{\today}

\begin{abstract}
We calculate the $f$-mode frequency and tidal overlap of quark stars using the full general relativity method.
We verify the universal relations obtained from conventional neutron stars in the case of quark stars and explore the cases with different values of parameters of the quark star equation of state.
Since quark stars have significantly smaller radii compared to neutron stars in the low mass range, the relation between the tidal defomability and $f$-mode frequency times radius is different for neutron stars and quark stars.
This difference has an impact on dynamical tide, which is the lowest-order effect we know of that can distinguish quark stars and neutron stars from the gravitational wave during the inspiral phase.
We calculate the tidal dephasing caused by this effect in the post-Newtonian method and find that it can not be detected even by the next-generation gravitational wave detectors.
\end{abstract}

\maketitle

\section{Introduction}

The state of cold matter at supranuclear densities is one of the most compelling problems in nuclear physics and astrophysics.
Although reliable calculations are feasible at ultra-high densities ($\agt10^2$ times the nuclear saturation density) which likely are irrelevant to any astrophysical systems, ab initio
calculations become notoriously difficult once the density drops below a few tens of times the nuclear saturation value, because quantum chromodynamics is strongly non-perturbative there.
Neutron stars (NSs) offer a unique avenue for probing the equation of state (EoS) of matter in this regime, precisely where theory struggles most and inaccessible to laboratory experiments.
Phenomenology of them thusly constitutes the primary pathway to uncover the fundamental physics at this density regime (such as the appearance of exotic particles and the feature of possible phase transition), and to guide the development of nuclear theory.

In particular, the binary neutron star (BNS) merger event GW170817~\cite{2017PhRvL.119p1101A} significantly improves our knowledge of EoS of cold matter, based on the tidal deformability inferred from the gravitational wave (GW) signal in the inspiral phase~\cite{2018PhRvL.120q2703A,2018ApJ...862...98Z} as well as the accompanied postmerger electromagnetic counterparts~\cite{LIGOScientific:2017ync}, which cover a gamma-ray burst, X-ray afterglow, kilonova, and late-time optical emission.
The launch of these electromagnetic emissions indicates a hypermassive NS was momentarily formed after the merger, which sheds certain light on the maximum mass of a cold non-rotating NSs~\cite{2017ApJ...850L..19M,2017PhRvD..96l3012S,2018ApJ...852L..25R,2018PhRvD..97b1501R,PhysRevD.100.023015}. 

During the late inspiral stage of a BNS merger, the influence of EoS are mainly imprinted in the GW signal via finite size effects.
The leading order contribution to the GW phasing from these effects is described by the tidal deformability~\cite{2008PhRvD..77b1502F,Hinderer:2007mb,Hinderer:2009ca,Damour:2009vw}.
The next order of (conservative) tidal effects stem from the dynamical tide to be interpreted as the frequency-dependent part of tidal response.
A popular way to model dynamical tides is through decomposing the excited perturbations as a sum of oscillation modes of NSs~\cite{Press:1977ApJ,Andersson:2019ahb,HegadeKR:2024agt}.
This interpretation has been extensively analyzed and incorporated into the SEOBNRT waveform models~\cite{PhysRevLett.116.181101,2016PhRvD..94j4028S,Steinhoff:2021dsn,Haberland:2025luz} and the FMTIDAL model~\cite{2019PhRvD.100b1501S}, while we note that there are other state-of-the-art analytic waveform models that do not count on the mode-sum approach to solve for the dynamical patch of tidal dephasing known as TEOB family~\cite{Damour:2009wj,Bini:2012gu,Bernuzzi:2014owa,Nagar:2018zoe,Akcay:2018yyh,Gamba:2023mww}.
In addition, phenomenological models calibrated from numerical relativity waveforms are also available and powerful in fast production of precise waveform templates for data analysis such as NRTidal~\cite{2017PhRvD..96l1501D,Dietrich:2018uni,2019PhRvD.100d4003D,Abac:2023ujg}, the one proposed in~\cite{2018PhRvD..97d4044K}, and PhenomGSF~\cite{Williams:2024twp}.
The signature of this dynamical tides could be relevant for future detectors especially for the next-generation infrastructures~\cite{Pratten:2019sed,Ma:2020oni,Pratten:2021pro,Williams:2022vct}; namely, Einstein Telescope (ET;~\cite{Punturo:2010zz,Hild:2010id,Maggiore:2019uih,Branchesi:2023mws,Abac:2025saz}) and Cosmic Explorer (CE;~\cite{LIGOScientific:2016wof,Reitze:2019iox,Evans:2021gyd}).

The effects of the static and dynamical tides are already minor compared to the contribution from bulk properties such as the chirp mass, it is naturally difficult to disentangle contributions from them in a given GW data. 
Nevertheless, considering both effects would enables much tighter constraints on the EoS, as the dynamical tide carries additional information beyond the tidal deformability; for instance, it reflects the radius of stars, which differs substantially between NSs and QSs. 
Realizing this potential requires precise modeling of tidal dephasing throughout the inspiral, which can only be possible based on accurate knowledge of the stationary and dynamical aspects of tidal effects.
From the perspective of decomposing dynamical tides into oscillation modes (see~\cite{HegadeKR:2025qwj,Andersson:2025iyd} for the recent elucidation), we then need a prudent determination of stellar oscillation spectrum.

There are many classes of quasi-normal modes for compact stars~\cite{McDermott:1985ApJ,McDermott:1988ApJ}, such as acoustic modes, gravity (g-)modes for interior that is stratified ~\cite{McDermott:1983ApJ,Reisenegger:1992ApJ,Lai:1993di,Kuan:2022bhu,Gittins:2024oeh,2024ApJ...964...31M} or possess discontinuity~\cite{Finn:1987MNRAS,Kruger:2014pva}, and interfacial and shear modes if a crust is present \cite{Schumaker:1983MNRAS,Finn:1990MNRAS,Yoshida:2002vd,VasquezFlores:2017tkp,Sotani:2023ypt,Sotani:2024hhi,Gao:2025aqo} (see the review~\cite{Kokkotas:1999bd} for more comprehensive categories).
In the acoustic class, the fundamental mode, which has no radial node, is dubbed as $f$-mode, while its overtones are collectively named as pressure (p-)modes.
The ability of exciting each oscillation modes by tidal field is characterized by the so-called tidal overlap quantity, which essentially describes how similar it is between the oscillation mode and the tidal field, in terms of matter distribution and frequency~\cite{Press:1977ApJ,Reisenegger:1994ApJ}.
Among these modes, the tidally driven $f$-mode is expected to have the most significant impact on the inspiral GW because the coupling strength to the tidal field of it dominates over those of other modes~\cite{Lai:1993di,Shibata:1993qc,Kokkotas:1995xe,Ma:2020rak,Passamonti:2020fur,2021MNRAS.506.2985K,Suvorov:2024cff,Gao:2025aqo}.

Previous studies have identified several nearly EoS-independent (universal) relations linking the tidal deformability of NSs to other properties such as the moment of inertia and quadrupole moment, compactness (the ratio of stellar mass to radius), and the frequencies of certain oscillation modes~\cite{2013Sci...341..365Y,2013PhRvD..88b3009Y,2020PhRvD.101l4006J}.
Most of these relations also hold for quark stars (QSs;~\cite{1970PThPh..44..291I,1986ApJ...310..261A}) albeit QSs have very different mass-radius relations\cite{Pretel:2024pem}.
Although the compactness relation differs for QSs\cite{2017PhR...681....1Y}, this difference does not directly affect the gravitational-wave waveform.
Universal relations can reduce the number of tidal parameters to be estimated, improving the accuracy of parameter extraction from GW signals~\cite{2017CQGra..34a5006Y}.
However, the fact that these relations hold for both NSs and QSs implies that static tidal effects in the inspiral are insufficient to distinguish the two, even if their radii and compactness differ significantly at fixed mass and tidal deformability.

This degeneracy may be lifted when dynamical tide effects are included in the analysis: QSs possess a finite surface density and a markedly different internal density profile compared to NSs, and, as a consequence, the universal relations concerning about stellar oscillation spectrum are expected to exhibit different behavior.
To the leading order, three variables  prove sufficient to specify the Hamiltonian governing the leading order dynamical tidal effects.
Namely, $RQ_f/M$, $\omega_fR$ and the tidal defomability for NSs, where $M$ and $R$ are the mass and radius of the star, $Q_f$ is the tidal overlap, and $\omega_f$ is the $f$-mode frequency.
It has been shown in Ref.~\cite{2022PhRvD.106f4052K} that universal relations exist between them for NSs.
Given the considerable difference in QS' radii, we aim to assess the extent to which universal relations involving the radius are violated for QSs.
We then estimate the impacts of this violation on the inspiral GW signals for canonical cases. 
It is found that the relation between $\omega_fR$ and the tidal deformability has an about 20\% violation for QSs.
While the associated signatures on the waveforms are not discernible to current detector, they may be picked up by ET and CE.


The paper is organized as following:
the calculation of the $f$-mode frequencies and the tidal overlap in full general relativity is introduced in \cref{sec:osc}.
In \cref{sec:uni}, we refit the universal relations from the data of NSs and QSs.
In \cref{sec:dep}, we calculate the GW delta phase of QSs and NSs, and show their difference caused by the difference in the dynamical tide.
In \cref{sec:de}, we consider detectability of this difference in phase between NSs and QSs by calculating the mismatch and signal-to-noise ratio(SNR).
We present our conclusions in \cref{sec:dc}, where we also discuss the effect of spin of the star on the tidal dephasing.

\section{oscillation frequencies and tidal overlap \label{sec:osc}}
In this work, the oscillation of the star is calculated in full general relativity, which means both the linear perturbations of the matter and those of the metric are considered.
For the ease of solving the quasi-normal mode spectrum, we adopt the Regge-Wheeler gauge \cite{1967ApJ...149..591T} to describe the perturbations as
\begin{align}
    \dd s^2 &= - e^{\nu(r)} \left(1+r^\ell H_0(r)e^{i\omega t} Y_{\ell m}(\phi,\theta) \right)c^2 \dd t^2\nn
    &+ e^{\lambda(r)} \left(1-r^\ell H_0(r)e^{i\omega t} Y_{\ell m}(\phi,\theta)\right)\dd r^2 \nn
    &+ \left(1-r^\ell K(r)e^{i\omega t}Y_{\ell m}(\phi,\theta)\right)r^2 \dd\Omega^2\nn
    &- 2i\omega r^{\ell+1}H_1(r)e^{i\omega t}Y_{\ell m}(\phi,\theta) \dd t \dd r\,,
\end{align}
where $H_0$, $H_1$, and $K$ are the metric perturbation functions. 
The system's dependence on the solid angle is described by the spherical harmonics, and it modulates with time in a harmonic manner dictated by the oscillation frequency $\omega$ ($\sim e^{i\omega t}$). 
The other terms depend only on the radius.
It is worthwhile to note that these perturbations describe even-parity modes, to which we restrict ourselves in the present study since odd-parity ones are (i) largely immune to tidal force and no sizable excitation would be resulted, and (ii) not featuring noticeable variation in the mass density and thus are not effective GW emitters.

The Lagrangian displacement vector is used to describe the perturbation of fluid.
The dependence of the components in each direction is expressed in terms of spherical harmonics as~\cite{1983ApJS...53...73L,1985ApJ...292...12D}
\begin{eqnarray}
\xi^r &=& r^{\ell-1}e^{-{\lambda}/{2}}W(r) Y^\ell_m e^{i\omega t}\,,\nonumber\\
\xi^\theta &=& -r^{\ell-2} V(r) \partial_\theta Y_m^\ell e^{i\omega t}\,,\\
\xi^\phi &=& -\frac{r^{\ell-2}}{ \sin^{2}\theta} V(r)\partial_\phi Y_m^\ell e^{i\omega t}\,,\nonumber
\end{eqnarray}
where $W$ and $V$ are the fluid displacement functions, which depend only on the radius, similar to the metric perturbation functions. 
For the convenience of calculations, we introduce $X$ defined through~\cite{1983ApJS...53...73L,1985ApJ...292...12D}
\begin{eqnarray}
V=\left [ \frac{X}{\varepsilon+p}+\frac{P'}{r}e^{-\lambda/2}W-e^{\nu/2}\frac{H_0}{2}\right ]\frac{e^{\nu/2}}{\omega^2} \label{eq:V_def_in}.
\end{eqnarray}
From the (0,1) component of the Einstein equation, we have another algebraic relation,
\begin{eqnarray}
H_0&=&\left\{8\pi r^2 e^{-\nu/2}X-\left[(n+1){Q}-\omega^2r^2 e^{-(\nu+\lambda)}\right]H_1 \nonumber \right. \\
&+&\left. \left[n-\omega^2 r^2 e^{-\nu}- {Q}({Q}e^{\lambda}-1)  \right] K \right\} (2b+n+{Q})^{-1}
\end{eqnarray}
where $n=(\ell-1)(\ell+2)/2$, $Q=b+4\pi G r^2 p/c^4$ and $b=Gm/(rc^2)$ with $m(r)$ is the total mass for the matter enclosed within the radius $r$. 

From the other components of the Einstein equation, the first-order, homogeneous linear differential equations are derived~\cite{1985ApJ...292...12D},
\begin{eqnarray}
\frac{\dd H_1}{\dd r}&=&- \frac{1}{r} \left[\ell+1+2 b e^\lambda+4\pi r^2 e^\lambda(p-\varepsilon)\right]H_1\nonumber\\
&&\quad\quad +\frac{e^\lambda}{r}\left[H_0+K-16\pi(\varepsilon+p)V\right],\nonumber
\\
\frac{\dd K}{\dd r}&=&\frac{1}{r}H_0+\frac{1}{r}(n+1)H_1\nonumber\\
&&+\frac{1}{r}\left[e^\lambda Q-\ell-1\right]K-\frac{8\pi}{r}(\varepsilon+p)e^{\lambda/2}W,\nonumber
\\
\frac{\dd W}{\dd r}&=&-\frac{\ell+1}{r}\left[W+\ell e^{\lambda/2}V\right]\label{eq:ODE_DL3}\\
&&+r e^{\lambda/2}\left[\frac{X}{ (\varepsilon+p) c_s^2 }e^{-\nu/2}+\frac{H_0}{2}+K\right],\nonumber
\end{eqnarray}
\begin{eqnarray}
    \frac{\dd X}{\dd r}&=&-\frac{\ell}{r} X+\frac{\varepsilon+p}{2r}e^{\nu/2}\left\{\left(3e^\lambda Q-1\right)K
    -\frac{4(n+1) e^\lambda Q}{r^2}V\right.\nn
    &+&(1-e^\lambda  Q)H_0+(r^2\omega^2e^{-\nu}+n+1)H_1+\left[-2\omega^2 e^{\lambda/2-\nu}\right.\nn
    &-&\left. \left.8\pi(\varepsilon+p)e^{\lambda/2}+r^2\frac{\dd}{\dd r} \left(\frac{e^{-\lambda/2}}{r^2} \frac{\dd\nu}{\dd r} \right)\right]W \right\}\,,
\end{eqnarray}
where $c_s$ is the sound speed in neutron or quark matter, neglecting the change of chemical composition, and $c_s^2=\dd p/\dd\varepsilon$ is assumed~\cite{2022PhRvD.106l3002Z}.
In this paper, we focus on modes with $\ell=2$ as these modes are subject to the leading order tidal interaction that stems from the quadrupole moment of tidal field (e.g.,~\cite{1977A&A....57..383Z}).

Quasi-normal modes are solutions to the linearized equations that obey certain and boundary conditions at the stellar center and surface, and have a pure outgoing behavior at spatial infinity.
In particular, considering the solution of the even parity, the first-order derivatives of $H_1$, $K$, $W$ and $X$ should be zero at $r=0$, yielding the algebraic relations:
\begin{align}
    X(0)&=(\varepsilon_0+p_0)e^{\nu_0/2}\,\times \nonumber\\
     &\quad \left \{ \left[ \frac{4\pi}{3}(\varepsilon_0+3p_0)-\frac{\omega^2}{\ell} e^{-\nu_0}\right]W(0)
    + \frac{K(0)}{2}\right \}, \label{eq:BC_H1}\\
    H_1(0)&=\frac{\ell K(0)+8\pi(\varepsilon_0+p_0)W(0)}{n+1}.\nonumber
\end{align}
They also need to satisfy the condition that the Lagrangian variation of pressure at the surface of the stars $\Delta p = -r^{\ell}e^{-{\nu}/{2}}X Y_m^\ell e^{i\omega t}$ vanishes [i.e., $X(R)=0$].
In addition, we fix $W(0)=1$ without loss of generality.
In order to satisfy the outer boundary condition, two trial solution are solved with $K_1(0)=\varepsilon_0+p_0$ and $K_2(0) = -(\varepsilon_0+p_0)$ and then calculate $K(0)$ via 
\begin{align}
    K(0) = \frac{K_1(0)X_2(R)-K_2(0)X_1(R)}{X_2(R)-X_1(R)},
\end{align}
where the different subscript represents the different trial solution~\cite{2022PhRvD.106l3002Z}.

To find the frequencies of the quasi-normal modes, one also need to solve the metric perturbations outside the star.
There is no matter outside the star (i.e., $W=X=0$), so the Zerilli equation for the perturbations of the metric ($K$ and $H_{1}$),
\begin{eqnarray}
    {\dd^2Z}/{\dd r^{*2}}&=&\left[ V_Z(r)-\omega^2 \right]Z\,, \label{eq:zerilli}
\end{eqnarray}
is used to calculate the exterior solution~\cite{1970PhysRevLett.24.737,1971ApJ...166..197F,1975RSPSA.344..441C}, where the effective potential $V_Z(r)$ is
\begin{eqnarray}
\hspace*{-.5cm}V_Z(r)=(1-2 {b} ) \frac{2n^2(n+1)+6n^2 {b} +18n {b} ^2+18 {b} ^3}{r^2(n+3 {b} )^2}\,,
\end{eqnarray}
and the tortoise coordinate $r^*=r+2m \ln(r/2m-1)$ is introduced.
The relations between the new variables $r^*$, $Z(r^*)$ and $\dd Z(r^*)/\dd r$ and the original ones ($r$, $H_1$ and $K$) are shown below,
\begin{eqnarray}
\begin{pmatrix}
K(r)\\H_1(r)
\end{pmatrix}
&=&
\begin{pmatrix}
g(r) & 1\\
h(r) & k(r)
\end{pmatrix}
\begin{pmatrix}
{Z(r^*)/r}\\ {\dd Z(r^*)/\dd r^*}
\end{pmatrix},\nonumber \\
g(r)&=&\frac{n(n+1)+3n {b} +6 {b} ^2}{(n+3 {b} )}, \\
h(r)&=&\frac{(n-3n {b} -3 {b} ^2)}{(1-2 {b} )(n+3 {b} )},\nonumber \\
k(r)&\equiv&\frac{\dd r^*}{\dd r}=\frac{1}{1-2 {b} }.\nonumber
\end{eqnarray}
A solution to the Zerilli equation has an asymptotic behavior,
\begin{equation}
\begin{aligned}
\label{Zasympt}
Z &\rightarrow
\left\{
\alpha - \frac{n+1}{\omega}\frac{\beta}{r} - \frac{1}{2\omega^2}\left[n(n+1)\alpha \right.\right. \\
&\left.\left. \quad - 3M\omega\left(1 + \frac{2}{n}\right)\beta\right]\frac{1}{r^2} + \dots \right\} \cos(\omega r_*) \\
&\quad - \left\{\beta + \frac{n+1}{\omega}\frac{\alpha}{r} - \frac{1}{2\omega^2}\left[n(n+1)\beta \right.\right. \\
&\left.\left. \quad + 3M\omega\left(1 + \frac{2}{n}\right)\alpha\right]\frac{1}{r^2} + \dots \right\} \sin(\omega r_*),
\end{aligned}
\end{equation}
as approaching infinity~\cite{1991RSPSA.432..247C,2003MNRAS.338..389M}, where $\alpha$ and $\beta$ represent the real and imaginary part of the amplitude of the standing gravitational wave at infinity.
These two parameters are obtained by matching $Z$ and $\dd Z/\dd r$ of the asymptotic solution with the numerical one.

In practice, we first search though the real axis to locate the ``neighborhood'' of the oscillating frequency and then figure out the imaginary part of the eigenfrequency from a fitting procedure according to Ref.~\cite{1991RSPSA.432..247C}. 
The relation between the square of the amplitude $\alpha^2 +\beta^2$ and the frequency $\omega$ can be related via
\begin{equation}
    \alpha^2 +\beta^2 = \rm const. \left[(\omega-\omega_0)^2+ \omega_i^2\right]\,.
\end{equation}
Fitting the numerical results to this behavior thus allows for determining the real ($\omega_0$) and imaginary ($\omega_i$) parts of the eigenfrequency.
This method has been proven to be equivalent to solving the eigen-problem through a search on complex plane for $\omega_{\rm i}\ll\omega_0$, satisfied by the modes of interest here~\cite{1991RSPSA.434..635C}.
Regarding the determination of $\omega_0$, this connection can be understood by analogizing the stellar oscillations to the forced vibrations with GWs acting as the driving force.
When a driving frequency approaches $\omega_0$ of a given mode, the fluid will undergo a `resonance', leading to large-amplitude motion~\cite{Thorne:1969ApJ}.
As a result, if the fluid amplitude is held fixed at the center [say $W(0)=1$], the GW flux reaches a minimum at resonance.

\begin{figure}
    \includegraphics[width=0.45\textwidth]{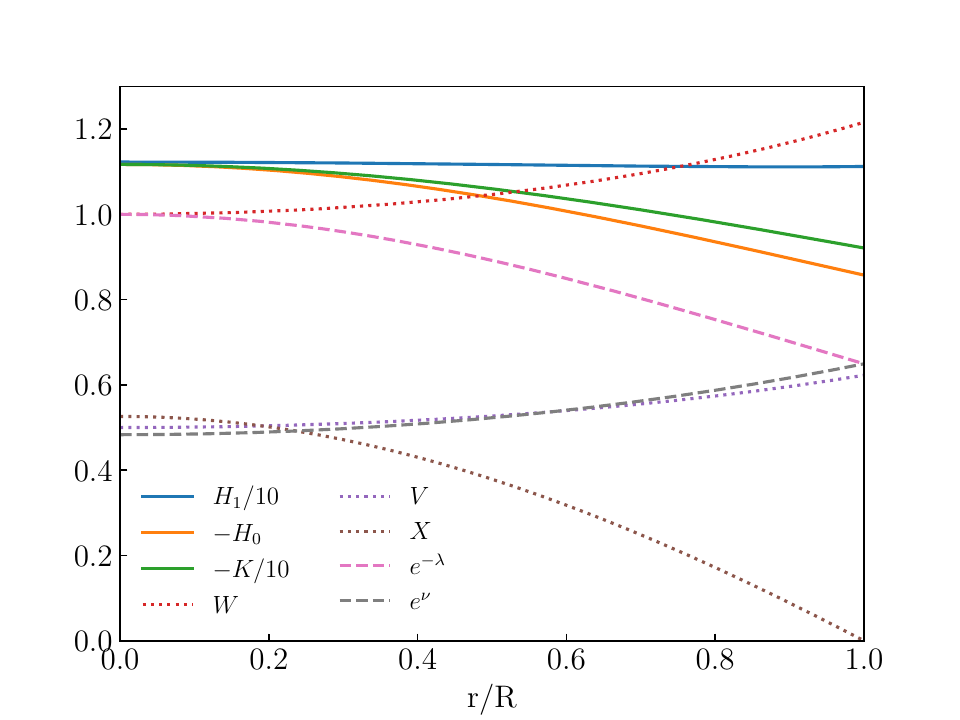}
    \caption{\label{fig:expqs} Metric perturbations (solid lines), fluid displacement functions (dotted lines), and metric functions of the static background (dashed lines) as functions of radius inside a 1.4~$M_{\odot}$ QS computed with a MIT bag model with $B = 47.202~\mathrm{MeV\,fm}^{-3}$.
    $H_0$, $H_1$ and $K$ are in units of $\varepsilon_s=152.26$ MeV fm$^{-3}$, $X$ is in units of $\varepsilon_s^2$, and $W$, $V$, $\nu$ and $\lambda$ are dimensionless. 
    The central pressure is $p_c=85$ MeV fm$^{-3}$, the radius is $R=11.72$~km, and the corresponding $f$-mode frequency is $\omega=(9.835\times 10^3+6.386~i)$~Hz, and the imaginary component of the frequency gives a damping time of 0.16~s. 
    }
\end{figure}

With the system of equations completed, boundary conditions imposed, and the numerical algorithm specified, the eigenproblem can now be readily solved.
As an example, we depict the $f$-mode for a $1.4\,M_\odot$ QS pertaining to the MIT bag model $P = \frac{1}{3}(\varepsilon-4B)$, 
with a bag constant of $B = 47.202~\mathrm{MeV\,fm}^{-3}$ in \cref{fig:expqs}.
We see that $X$ vanishes at the stellar surface, reflecting the boundary condition.

After obtaining the oscillation frequency, another important aspect to consider is how easily a certain mode can be excited by the tidal interaction in the late inspiral phase. 
The quantity used to describe this property is the tidal overlap~\cite{1999ApJ...515..414Y,2021MNRAS.506.2985K,Counsell:2024pua} (see also \cite{Gao:2025aqo} for another flavor of definition while we note that both yield nearly identical predictions for the mode’s tidal excitation),
\begin{equation}\label{eq:rel Q}
    Q_\alpha=\frac{1}{MR^2}\int\sqrt{-g}e^{-\nu}\dd^3x~(p+\varepsilon)~\bar{\xi}_\alpha^{*\mu}~\nabla_\mu\left(r^2Y_{2m}\right)\ ,
\end{equation}
where $g$ is the determinant of the metric, $\bar{\xi}^\mu_\alpha$ denotes the normalized eigenfunction that satisfies the orthogonality
\begin{align}
    \int\sqrt{-g}e^{-\nu}\dd^3x~(p+\varepsilon)~\xi^{*\mu}_\alpha~\xi_{\mu \alpha'}=\delta_{\alpha \alpha'} MR^2\,.
\end{align}
In the context of quasi-circular and non-processing orbits, it is sufficient to consider modes with $m=0$ and $\pm2$ for the tidal excitation problem since the leading order tidal field is of quadrupole and the ones with $m=\pm1$ are decoupled from the tidal forcing.

\begin{table*}
\caption{Properties of QSs and NSs with nearly the same mass $M$ and tidal deformability $\Lambda$.
The first column shows the stellar mass $M$ ($M_\odot$).
For both QSs (second to fifth columns) and NSs (sixth to ninth columns), the tabulated quantities are: radius $R$ (second and sixth columns), tidal deformability $\Lambda$ (third and seventh columns), $f$-mode frequency $f$ (fourth and eighth columns), and tidal overlap $Q_f$ (fifth and ninth columns).
For NSs, the EoS named RGSK(272) is used. 
For QSs, the MIT bag model with different B are used to guarantee they have the same static tides with NSs at the same mass. 
Specifically, for the mass from top to bottom row, we set $B =45.399,\,46.472,\,47.481,\,48.476,\,49.427,\,50.350,\,51.217\,\, \rm MeV\, fm^{-3}$, respectively. }
\label{table:data}
\begin{ruledtabular}
\begin{tabular}{ccccccccc}
\textbf{M ($M_\odot$)} & \multicolumn{4}{c}{\textbf{quark stars}} & \multicolumn{4}{c}{\textbf{neutron stars}} \\
\cline{2-5} \cline{6-9}
 & \( R \) (km) & \( \Lambda \) & \( f \) (Hz) & \( Q_f \) & \( R \) (km) & \( \Lambda \) & \( f \) (Hz) & \( Q_f \) \\
\hline
1.10 & 11.214 & 2709.83 & 1490.24 & 0.771 & 13.460 & 2712.48 & 1486.20 & 0.637 \\
1.15 & 11.260 & 2097.75 & 1515.82 & 0.775 & 13.435 & 2101.54 & 1511.59 & 0.644 \\
1.20 & 11.304 & 1636.70 & 1540.68 & 0.780 & 13.410 & 1638.54 & 1536.84 & 0.651 \\
1.25 & 11.343 & 1282.87 & 1565.76 & 0.784 & 13.383 & 1284.74 & 1561.93 & 0.658 \\
1.30 & 11.378 & 1010.73 & 1590.63 & 0.789 & 13.354 & 1012.13 & 1587.00 & 0.666 \\
1.35 & 11.410 & 799.21 & 1615.67 & 0.793 & 13.323 & 800.57 & 1611.99 & 0.673 \\
1.40 & 11.439 & 634.75 & 1640.44 & 0.798 & 13.290 & 635.44 & 1637.16 & 0.680 \\
\end{tabular}
\end{ruledtabular}
\end{table*}

\section{universal relations\label{sec:uni}}
There exist certain relations between the properties of stars, known as universal relations, which hold for all plausible EoS.
Some of them prove to be also applicable to QSs~\cite{1970PThPh..44..291I,1986ApJ...310..261A,Pretel:2024pem}.
However, the non-zero surface density and the different density distribution of QSs cause some relations to differ when applied to QSs and NSs.
In particular, QS and NS could be very different in radius for given values of mass and tidal deformability, potentially leading to different dynamical tide effects. 
\begin{figure}[b]
\includegraphics[width=0.45\textwidth]{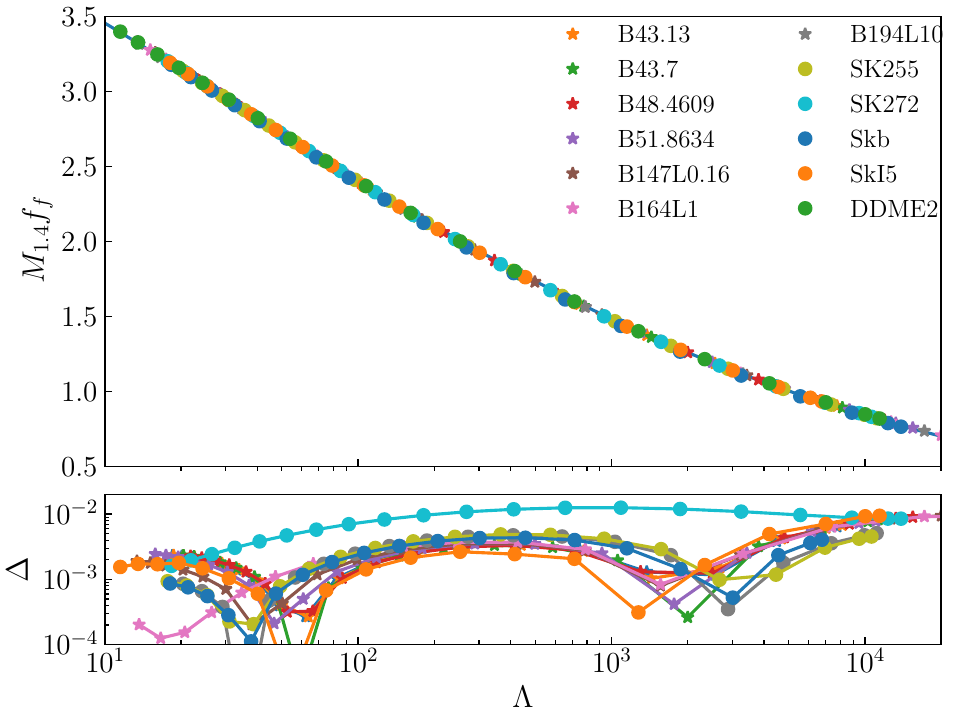}
\caption{\label{fig:ol} Top panel: universal relation between the $f$-mode frequency times $M_{1.4}$ and the tidal defomability. The star-shaped points represent our data for QSs, some of them are modeled in the MIT bag model with the different values of the effective bag parameter $B$, which reflect the surface density of QSs, as indicated in the legend. For other QSs, the EoS in Ref.\cite{2021PhRvD.103f3018Z} is used, all possible strongly interacting phases of strange quark matter and ud quark matter is considered in this model. The parameter $\bar{\lambda}$ characterizing the strength of the related strong interaction. In the legend, the values of this parameter are showed by the number after "L".
The circular points represent the results of NSs.
The blue solid line show the result in Ref.~\cite{PhysRevD.104.123002}. 
Bottom panel: relative deviation between our results and the relation discussed in Ref.~\cite{PhysRevD.104.123002} as a function of the tidal defomability.} 
\end{figure}

In this paper, we aim to search for the largest possible difference in dynamical tides of BQS and BNS mergers, and hence we need to chose NS and QS EOSs which produces largest differences in radius yet keeping mass and tidal deformability the same. 
{For this purpose, we adopt neutron-star EOSs DDME2\cite{2025arXiv250611492N}, aas well as several EOSs from the RG (Gulminelli–Raduta)\cite{PhysRevC.92.055803} unified family, each based on a different effective interaction: SK255 and SK272~\cite{PhysRevC.68.031304}, Skb\cite{KOHLER1976301}, SkI5\cite{REINHARD1995467}.
These EOSs are chosen such that a 1.4~$M_\odot$ NS has a relatively large radius of $\sim13$~km and a tidal deformability $\Lambda<800$, to be in line with GW170817~\cite{LIGOScientific:2018cki}.
For QS EOSs, We consider the following two models: one is the simplest MIT bag model, which describes the non-interacting quark matter; 
another one is proposed in Ref.~\cite{2021PhRvD.103f3018Z}, which describes interacting quark matter including perturbative QCD corrections and color superconductivity for both two flavor and three flavor quark matter.
After reparameterization and rescaling, these EOSs contain one more parameter,
\begin{eqnarray}
  \label{eq:zc}
  \bar{P} = \frac{1}{3}(\bar{\varepsilon}-1) + \frac{4\bar{\lambda}}{9\pi^2}(-1+\rm sgn(\lambda)\sqrt{1+\frac{3\pi^2}{\bar{\lambda}}(\bar{\varepsilon}-\frac{1}{4})}),
\end{eqnarray}
 here $\bar{P}=P/4B$,$\bar{\varepsilon}=\varepsilon/4B$ and $\bar{\lambda}$ is a parameter characterizing the strength of the related strong interaction~\cite{2021PhRvD.103f3018Z}.
 In this paper,we restrict our attention to the case $\lambda>0$.

The universal relations for QSs found in~\cite{PhysRevD.104.123002} are used to verify our results.
The blue solid line in the upper panel of Fig.\ref{fig:ol} is the fitting results in their paper, which is 
\begin{eqnarray}
  \label{eq:ol}
 M_{1.4} f_f &=& 4.2590-0.47874\log(\Lambda)-0.45353\log(\Lambda)^2 \nn
 &&+0.14439\log(\Lambda)^3-0.016194\log(\Lambda)^4 \nn
 &&+0.00064163\log(\Lambda)^5,  
\end{eqnarray}
where $M_{1.4} = M/(1.4 M_\odot)$ and for convenience of comparison with their results, kHz is used as the unit of frequency in this figure which differ from all other figures in this paper.
As shown in this figure, all our data points are close to their fitting curve.
In order to show the difference between our data and their result more clearly, the lower panel show the relative deviation calculated with
\begin{equation}
    \label{eq:rd}
    \Delta = \frac{|M_{1.4} f_f-(M_{1.4} f_f)_{fit}|}{M_{1.4} f_f}.
\end{equation}
As shown in the lower panel of \cref{fig:ol}, most of this difference are below one percent which is very small, showing a perfect agreement with their results.

Some universal relations have been found in the case of NSs and have not been validated with data including QSs.
The relations about the dynamical tides found in Ref.~\cite{2022PhRvD.106f4052K} are particularly important since they include the radius which differs significantly between QSs and NSs,
\begin{equation}\label{eq:def_AB}
\begin{aligned}
    &\mathcal{A}(\Lambda) = Q_fR/M, \\
    &\mathcal{B}(\Lambda) = \omega R.    
\end{aligned}
\end{equation}

\begin{figure}[b]
\includegraphics[width=\columnwidth]{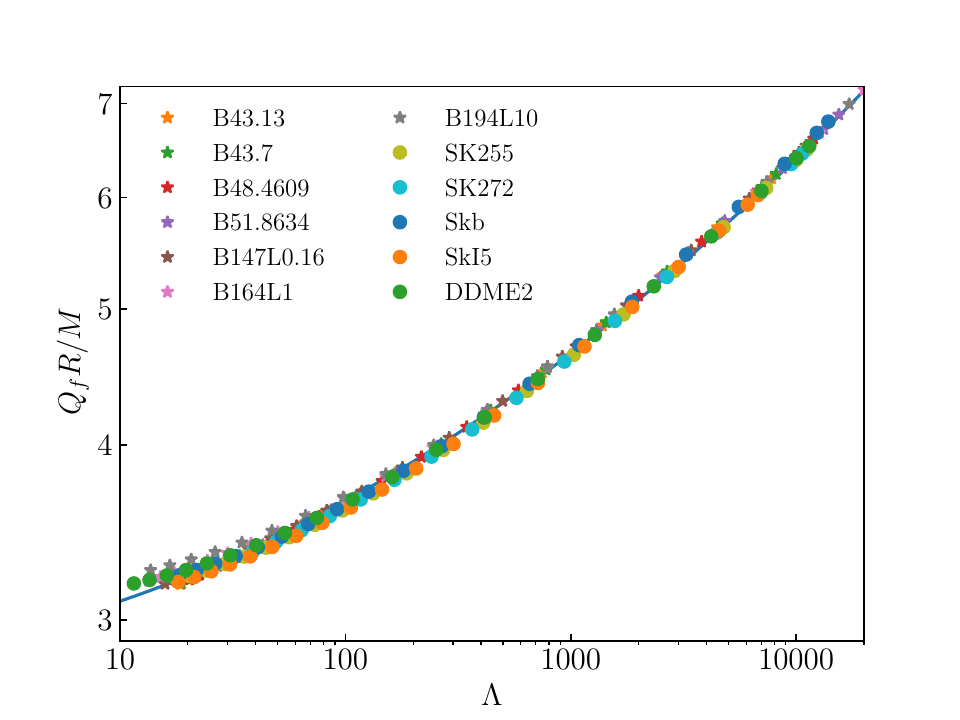}
\caption{\label{fig:al} Universal relation between $\mathcal{A} = Q_fR/M$ and the tidal defomability. 
The point shapes have the same meaning with Fig. \ref{fig:ol}.
The blue solid line shows our fitting result which is showed in Eq. (\ref{eq:al}).}
\end{figure}

\begin{figure}
    \includegraphics[width=\columnwidth]{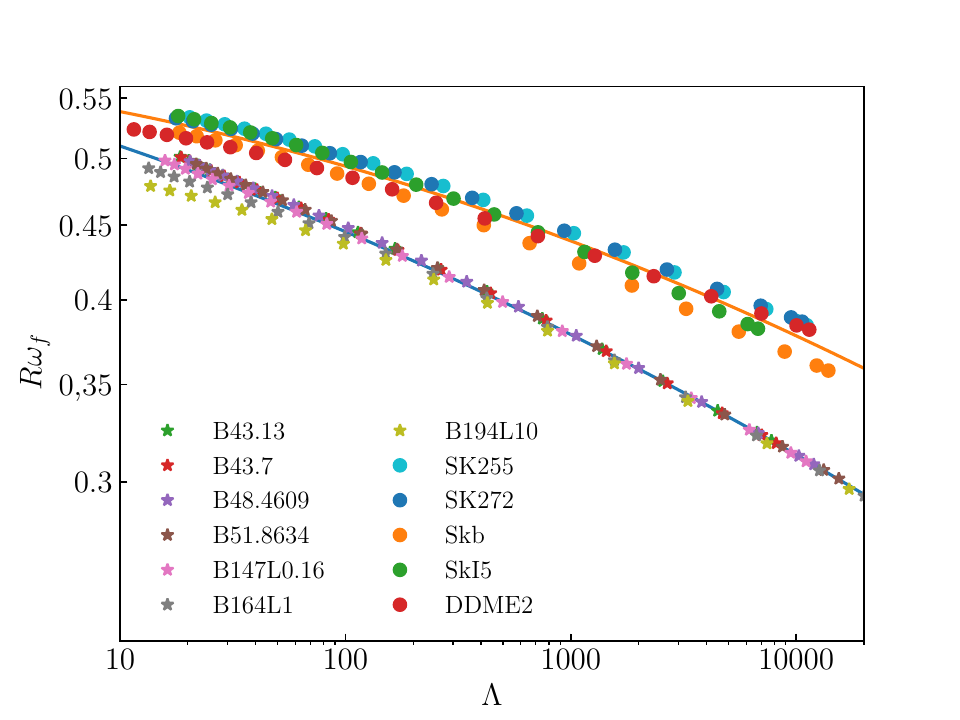}
    \caption{Universal relation between $\mathcal{B}(\Lambda) = \omega R$ and the tidal defomability. The point shapes have same the meaning with \cref{fig:ol}.
    The blue solid line shows our fitting result of NSs which is shown in \cref{eq:bln} and the orange solid line shows the fitting result of QSs which is shown in \cref{eq:blq}.
    \label{fig:bl}
    }
\end{figure}
Including the results of QSs, these relations are refitted in this paper.
\cref{fig:al} shows the universal relation of QSs between $\mathcal{A}$ and the tidal defomability of compact stars.
This figure illustrates that the data points of QSs and NSs lie almost on the same curve,
\begin{equation}
\label{eq:al}
    \log(\mathcal{A}) = 0.01687 \log(\Lambda)^2 + 0.02078 \log(\Lambda) + 0.453.
\end{equation}
In other words, this relation also applies to QSs.

\cref{fig:bl} shows that the universal relation of QSs between $\mathcal{B}$ and the tidal deformabilities is different with that of NSs. 
The fitting result for NSs is
\begin{equation}
\label{eq:bln}
    \log(\mathcal{B}) = -0.006918\log(\Lambda)^2 - 0.01665 \log(\Lambda) - 0.2453,
\end{equation}
while the result for QSs is 
\begin{equation}
\label{eq:blq}
    \log(\mathcal{B}) = -0.005922 \log(\Lambda)^2 - 0.04096 \log(\Lambda) - 0.2455\,.
\end{equation}
The deviation between the NS results and the fitted values is larger than that for QSs.
For QSs, this difference occurs primarily in regions with smaller tidal deformabilities, which means that these stars have larger mass and among EoSs with different values of $\bar{\lambda}$.

The reason of these results can be found in \cref{table:data}. 
NS EoSs with a large radius at 1.4~$M_\odot$ are used in this paper, which makes NSs have a large difference with QS one.
For QSs, the MIT bag model with different $B$ which make the star has nearly the same tidal defomability as NSs at the same mass is used.
The QSs and NSs have the similar $f$-mode frequencies, but QSs have a radius which is much smaller than that of NSs and also have a larger tidal overlap of the $f$-mode. 
The radius and tidal overlap of QSs are smaller than those of NSs by the same factor, resulting in a similar universal relation for $A$.
QSs have an $f$-mode frequency similar to that of NSs, but have a smaller radius, so the relation between $B$ and the tidal deformability for QSs and NSs is different.

\section{Tidal Dephasing\label{sec:dep}}
Dynamical tides imprint certain dephasing to the waveform which can be distinguished from the static tide effect when SNR is high enough and the signal can be detected up to merger~\cite{Pratten:2019sed,Ma:2020oni,Pratten:2021pro,Williams:2022vct}.
One can then infer the properties \cref{eq:def_AB} from the detected and analyzed signals.
As discussed above, this dynamical tide effect is modeled by the properties [cf.~\cref{eq:def_AB}], thus a detailed analysis and modeling can be combined to obtain some constrain on these parameters.
The obtained information can be further leveraged to infer indirect observables such as the stellar radius when combined with the universal relations \cref{eq:al,eq:bln,eq:blq}.
In addition, the dependence of ${\cal B}$ on $\Lambda$ differs between NSs and QSs.
These differences influence the tidal dephasing produced by the dynamical tide, and our goal is to quantify the magnitude of this effect and evaluate whether it could provide a clear, detectable signature of the stars’ true nature.



The post Newtonian method is used to calculate the tidal dephasing~\cite{2022PhRvD.106f4052K,Kuan:2023qxo}.
In this method, the motion of the binary system is obtained by solving the Hamilton's equations with the Hamiltonian~\cite{Alexander:1987zz,Reisenegger:1994ApJ},
\begin{align}\label{eq:ham}
	H=H_{\text{orb}} + H_{\text{reac}} + H^{T} ,
\end{align}
where the first two terms are the point-particle contributions: $H_{\text{orb}}$ is the Hamiltonian for the conservative motion for which we solve up to 3 post-Newtonian orders, and $H_{\text{reac}}$ is the Hamiltonian that considers gravitational backreaction~\cite{2021MNRAS.506.2985K}. 
Finite size effects including static and dynamical tides are grasped in the tidal Hamiltonian~\cite{2022PhRvD.106f4052K}
\begin{align}\label{eq:uni_H}
	H^{T}&=H_{\text{tid}}+H_{\text{osc}} \nonumber\\
    &=-\frac{2 {M}^2 M_{\text{comp}}}{a^3} {\cal A} q\cos(m\varphi_c) 
	+\left( \frac{p\bar{p}}{M R^2}+M q \bar{q} {\cal B}^2 \right)\,.
\end{align}
where $M$ and $M_{\text{comp}}$ are the mass of the primary star and companion star, and $a$ is the distance between them.  $\phi_{c}$ is the phase coordinate of the companion star, $q$ are the mode amplitudes, and $p$ are the canonical momenta associated with $q$.

\begin{figure}
    \includegraphics[width=\columnwidth]{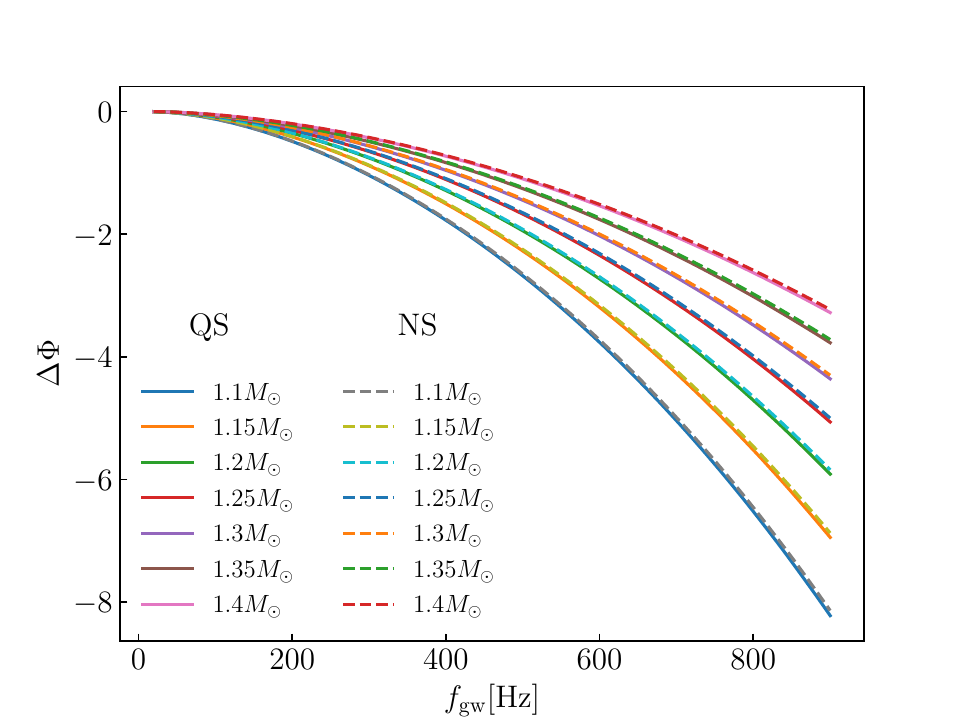}
    \caption{\label{fig:dp} Tidal dephasing of NSs and QSs as a function of GW frequency.
    Solid lines denote QSs and dashed lines denote NSs. The data are taken from \cref{table:data}, where QSs and NSs of the same mass have approximately equal tidal defomabilities.}
\end{figure}

We consider QSs that have tidal defomabilities similar to those of NSs, as listed in \cref{table:data}, so that the static tidal responses of QSs and NSs are comparable.
In addition, we restrict ourselves to equal-mass, non-spinning binary systems in the present study.
The GW phase is obtained by integrating over a frequency range $[f_{\rm min},\,f_{\rm max}]$,
\begin{align}\label{eq:dep}
    \Phi_{\rm tot}=\int^{f_{\rm max}}_{f_{\rm min}}\dd f_{\rm gw}\left(\frac{\partial \Phi}{\partial f_{\rm gw}} \right)
    =\Phi_{\rm pp}+\Delta\Phi^T\,,
\end{align}
where $\Phi_{\rm pp}$ denotes the point-particle contribution to GW phase, and $\Delta \Phi^T$ collectively represents the tidal dephasing including the both equilibrium and dynamical tides.
Since tidal effects are negligible during sub-hundred Hz~\cite{Damour:2012yf,Harry:2018hke} and the post-Newtonian approximation may become unreliable above 900~Hz, we choose $(f_{\rm min},\,f_{\rm max})=(20,\,900)$~Hz and plot the results in \cref{fig:dp}.
This result includes the contributions from both static and dynamical tides. 
The tidal dephasing is sufficiently large to be detectable by aLIGO even for the case of $1.1~M_\odot$: at $f_{\rm gw}=900$ Hz, we find $\Delta\Phi = -3.24$ rad for NSs and $\Delta\Phi = -3.28$ rad for QSs.
The static tide provides the dominant contribution to the overall dephasing, which leads to larger tidal dephasing for stars with smaller masses due to their higher.
However, by selecting similar tidal deformabilities for both types of stars as done here, its effect on the difference between QSs and NSs is effectively removed, allowing us to focus on the next-order contribution, namely the dynamical tide.

Due to the more uniform density distribution of QSs, QSs have larger tidal overlaps than NSs, which means that their $f$-modes are more easily excited; this makes their effects of dynamical tides larger than NSs.
The difference in tidal dephasing between NSs and QSs is small in this figure. \cref{fig:ddp} is included as a supplementary figure to illustrate this difference for NSs and QSs of similar mass and tidal deformability.
We see that the difference between NSs and QSs in tidal dephasing increases with decreasing mass since the difference in radius is larger at lower masses, leading to a larger difference in $\cal B$.
That said, the difference in $\cal B$ is still very small (only $0.0822$ at $f_{\rm gw} = 900$~Hz) even for the smallest mass discussed in this paper (i.e.,~$1.1 M_{\odot}$).
Meanwhile, it is worth noting that although the tidal deformabilities of the selected QSs and NSs are not exactly the same, their difference is only at the level of $10^{-3}$, which is much smaller than the difference in the tidal dephasing (at the level of $10^{-2}$). This indicates that the dominant factor of the tidal dephasing difference is the dynamical tides.

\begin{figure}
    \includegraphics[width=\columnwidth]{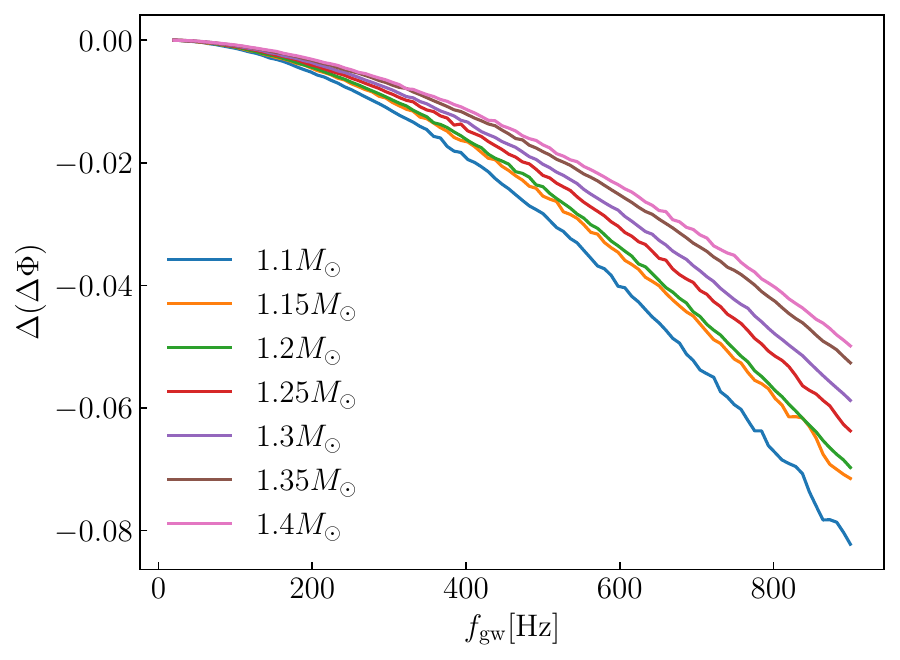}
    \caption{Difference in tidal dephasing between NSs and QSs of similar mass and tidal deformability as a function of GW frequency.}
    \label{fig:ddp} 
\end{figure}

Before we assess the detectability of this difference with current and future GW detectors in \cref{sec:de}, we briefly comment on the possible influence of spin.
To the leading order, the frequency shifts induced by spin is given as~\cite{Strohmayer:1991ApJ,Vavoulidis:2007cs,2021MNRAS.506.2985K},
\begin{align}\label{eq:domg}
    \delta \omega=-m\Omega (1-C_{nl}),
\end{align}
with
\begin{align}\label{eq:cnl}
    C_{nl}=&\frac{1}{MR^2}\int e^{-\nu/2}\dd^3x\left(p+\varepsilon\right)r^{2l}\nn
    &\left(-\bar{V}_{nl}W_{nl}-\bar{W}_{nl}V_{nl}+\bar{V}_{nl}V_{nl}e^{\lambda/2}\right).
\end{align}
We restrict to the $l=2$ $f$-modes, and find that the difference less than one percent in $1-C_{02}$ between the two kinds of EoS for all the mass from 1.1$~M_\odot$ to 1.4$~M_\odot$.
The negligible difference in spin modulation of the frequencies, and the presumably even smaller impact on the eigenfunctions~\cite{Lee:1996A&A}, suggest that this effect is unlikely to help distinguish NSs from QSs.

\section{Detectability analysis}
\label{sec:de}

The match (or faithfulness) between two waveforms is quantified by the measure~\cite{1996PhRvD..53.6749O},
\begin{align}\label{eq:match}
    \mathcal{F} = \frac{\langle h_1|h_2\rangle}{\sqrt{\langle h_1|h_1\rangle\langle h_2|h_2\rangle}}\,,
\end{align}
where $h_1$ and $h_2$ are the two waveforms in time domain, and the inner product between waveforms is weighted by the noise function of the detector $S_n(f)$.
In particular, the explicit form of the inner product reads
\begin{align}\label{eq:inn}
    \langle h_1|h_2\rangle:=4\mathfrak{R}\left[
    \int^{f_{max}}_{f_{min}}
    \frac{\tilde h_1(f)\tilde h_2^*(f)}{S_n(f)}
    \dd f\right]\,,
\end{align}
where $\tilde h_1(f)$ and $\tilde h_2(f)$ are the Fourier transforms of the time domain waveforms. 
We choose the same frequency range used to compute the GW phase, namely from $f_{\rm min} = 20$ Hz to $f_{\rm max} = 900$ Hz.
The match are equal to one for the two same waveforms, so we use the mismatch to quantify the difference between two waveforms,
\begin{align}\label{eq:mis}
    \mathcal{MM} = 1 -\mathcal{F}\,.
\end{align}
In the mismatch analysis, constant time and phase shifts are introduced to minimize the mismatch between the two waveforms.

For two waveforms which are distinguishable, they satisfy the criteria~\cite{Finn:1992wt,Cutler:1994ys,2008PhRvD..78l4020L},
  \begin{align}\label{eq:cri}
    \langle\delta \tilde h|\delta \tilde h\rangle>1,
\end{align}
where $\delta \tilde h$ is the Fourier transform of $\delta h$, and $\delta h$ is the difference between two waveforms in the time domain $\delta h(t) = h_1(t)-h_2(t)$.
The only distinction between the two signals is a small phase difference, the difference between $\langle \tilde h_1 | \tilde h_1 \rangle$ and $\langle \tilde h_2 | \tilde h_2 \rangle$ and is minor and neglected here.
In accordance, $\langle \delta \tilde h | \delta \tilde h \rangle$ can be reformulated as~\cite{2010PhRvD..82b4014M,2010PhRvD..82l4052H,2025arXiv250610530T}
  \begin{align}\label{eq:rec}
    \langle\delta \tilde h|\delta \tilde h\rangle&=2\langle \tilde h_1| \tilde h_1\rangle \left(1-\frac{\langle h_1|h_2\rangle}{\sqrt{\langle h_1|h_1\rangle\langle h_2|h_2\rangle}} \right)=:2\rho^2\mathcal{MM}(h_1,h_2)
\end{align}
 where $\rho=\sqrt{\langle \tilde h_1| \tilde h_1\rangle}\circeq\sqrt{\langle \tilde h_2| \tilde h_2\rangle}$ is the SNR of the waveforms under this GW detector.
 
 We calculate the mismatch between waveforms from the binary neutron star (BNS) and binary quark star (BQS) with the same mass and tidal deformability using pyCBC~\cite{pycbc2022}, which chooses the phases and time shifts to align the waveforms effectively.
 The waveforms are obtained by adding the tidal dephasing of NSs and QSs to a binary black hole baseline, here adopting TEOBResumS model~\cite{Nagar:2018zoe}, with the same mass.
 The estimate is based on the designed sensitivity curves of aLIGO \cite{2017CQGra..34d4001A}, ET \cite{2023JCAP...07..068B,2011CQGra..28i4013H} and CE \cite{2017CQGra..34d4001A,2023arXiv230613745E}.
 We assume that the GW signal is face-on to the Earth and the source is located at $D_L=50 ~\rm Mpc$.

We can see from \cref{fig:mis}, both the mismatch and $1/{\rm SNR}^2$ decrease with mass.
The reason for the former is that the difference of dephasing between QSs and NSs decreases with mass.
The latter is because the signal is weaker for a lower mass system, and the higher GW frequency makes it more difficult to be detected.
It can be read from the lower panel that the former effect is larger, making it easier to distinguish BNS and BQS with lower mass binaries.
The mismatches obtained from the noise function of the different detectors are similar, because the noise functions have a similar shape in the premerger phase.
Due to their higher sensitivity, the next-generation GW detectors (ET and CE) yield a SNR much higher than that achieved by aLIGO by about three orders of magnitude.

In particular, the maximum mismatch for aLIGO is found  at 1.1$~M_\odot$ as $8.950\times10^{-3}(\ll1)$.
This level of mismatch is buried in the noisy response of aLIGO, and hence the signal coming from BNS and BQS cannot be distinguished even if the dynamical tides are measured.
However, this difference could be marginally detected by the next-generation GW detector.
For systems with masses between 1.1 and 1.4~$M_\odot$, the mismatch can exceed the noise level of CE and thus could potentially be observable.
For ET, on the other hand, the mismatch is close to $1/\rm SNR^2$ for low masses, and 
${\cal M}\times{\rm SNR}^2$ only exceeds unity for the 1.1~$M_\odot$ case.
The detectability with ET is thus not very plausible though a close event could still be captured.

To strengthen our conclusion about the detectability, we also consider $(f_{\rm min},\,f_{\rm max}) = (30,\,900)~\rm Hz$.
It is found that the measurability does not suffer from the lack of low frequency signals due to a compensation effect,
i.e., the increase of mismatch, resulting from retaining only the high frequency portions where the difference between waveforms is larger, counteracts the lower SNR.
As a result, $\langle\delta \tilde h|\delta \tilde h\rangle$ remains almost the same as the results shown in \cref{fig:dfr}.

\begin{figure}
    \includegraphics[width=0.48\textwidth]{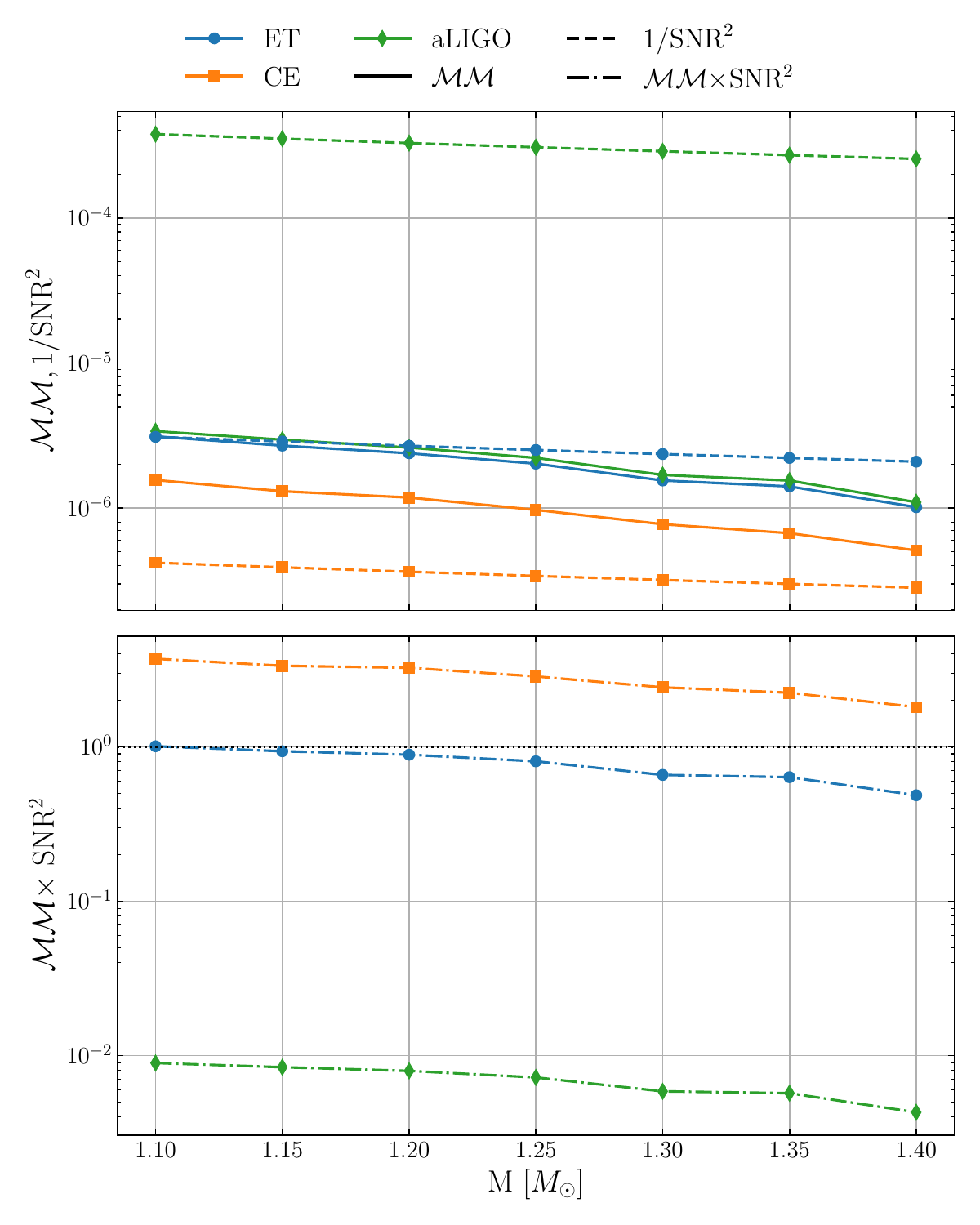}
    \caption{\label{fig:mis}Top panel: mismatch $\mathcal{M}$ of waveforms from BNSs and BQSs with the same tidal deformabilities as a function of  masses(the solid line). Also plotted are the corresponding $1/{\rm SNR}^2$(the dotted line). 
    The data obtained by the noise function of different GW detector are showed in different color and point shape. 
    Due to \cref{eq:cri,eq:rec}, if the mismatch is larger than $1/{\rm SNR}^2$ of the same detector which means the point in solid line is higher than the dotted line in the same color, the waveform can be detected by this detector.
    Bottom panel: $\langle\delta \tilde h|\delta \tilde h\rangle$ as a function of mass.}
\end{figure}

\begin{figure}
    \includegraphics[width=\columnwidth]{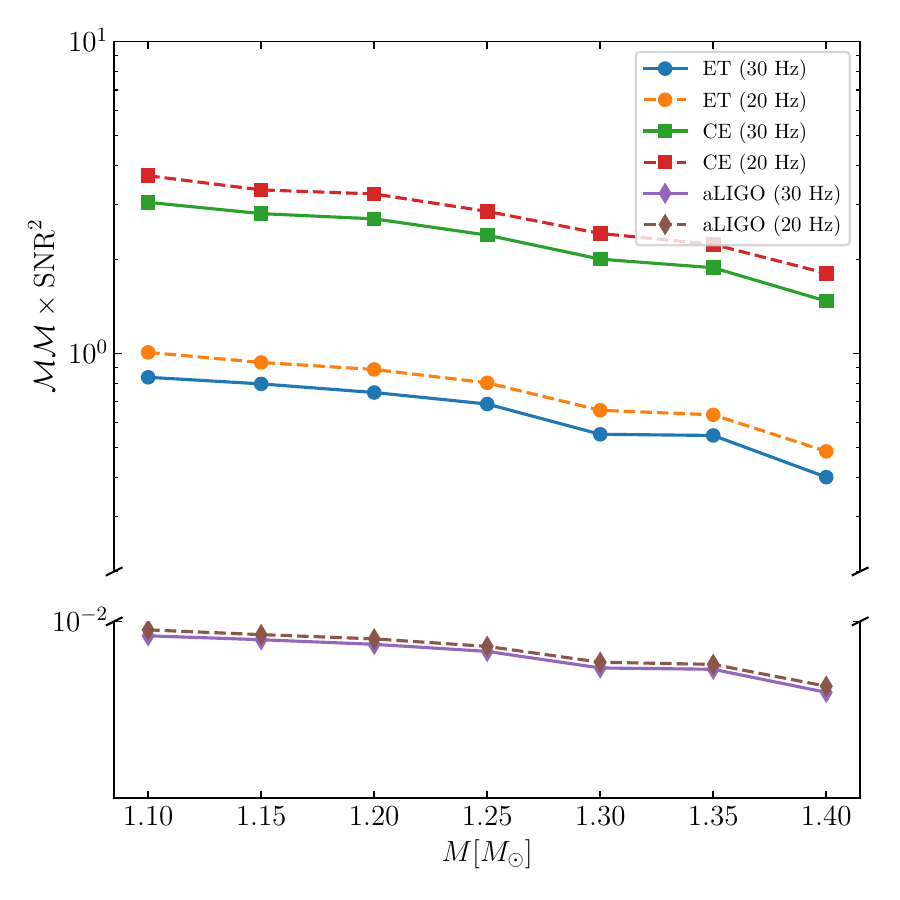}
    \caption{\label{fig:dfr}
    Difference of $\langle\delta \tilde h|\delta \tilde h\rangle$ with different frequency range (20--900~Hz and 30--900~Hz) as a function of mass.
    }
    \label{fig:dfr}
\end{figure}



\section{DISCUSSION AND CONCLUSION}
\label{sec:dc}
GW observations that distinguish between QSs and NSs can provide a crucial means to probe the ground state of nuclear matter under conditions of low temperature and high density.
In particular, differences in their internal structure suggest that finite-size effects in the waveforms could serve as a decisive diagnostic tool.
However, this goal is also bothered with the huge uncertainties in the EOS for either nature of the stars.
To access the feasibility of such distinction, we appeal to universal relations that connect the tidal parameters with bulk properties to minimize the ramifications of uncertainties of this kind.
This approach requires identifying a relation that differs between NSs and QSs. However, known relations for equilibrium tides take the same form for both types of stars.
We thus propose to solve this problem by considering the next order tidal effect, the dynamic tide, which is associated with the excitation of oscillation modes.

Dynamical tidal effects are primarily parameterized by the frequencies and tidal overlaps of $f$-mode.
Full general-relativity calculations for QSs and NSs show that QSs of the same mass and tidal defomability as NSs have similar $f$-mode frequencies but larger tidal overlaps.
The parameters that factor in the Hamiltonian governing the dynamical tides are $Q_fR/M$ and $\omega_f R$.
The formal is similar for both NSs and QSs, so the universal relation between $Q_fR/M$ and $\Lambda$ of QSs is the same as that for NSs.
However, due to the difference in radius and the similarity of their $f$-mode frequencies, the universal relation between $\omega_f R$ and $\Lambda$ differs by about 20\% between the two types of stars.

The discrepancy in the $\omega_f R$--$\Lambda$ relation is then translated to a waveform observable by computing the tidal dephasing of the dynamical tide using a PN method~[\cref{eq:uni_H}].
Due to the fact that the difference of density distribution makes the tidal overlap of QSs larger than that of NSs. The $f$-modes of QSs thus more efficiently absorb energy and angular momentum from the orbit to return  more significant tidal dephasing on waveforms.
Ultimately, this difference in the GW phase, integrated from 20 to 900 Hz, increases with lower masses. 
For an equal-mass, non-spinning binary with component masses of $1.1\,M_\odot$ -- the lowest mass considered in this paper -- the magnitude of this effect is merely 0.0822 radian.
We evaluate the mismatch between BNS and BQS waveforms across various GW detectors and provide the corresponding SNRs.
Combining our results with the previous conclusion that the static tides~\cite{2013Sci...341..365Y} and the GW cutoff frequency cannot distinguish them, it appears almost impossible to differentiate NSs and QSs from the GW signal during the inspiral phase by aLIGO.
Even for a $1.1,M_\odot$ case, detecting this difference with aLIGO would require an exceptionally loud event with an SNR exceeding 543.
However, the difference in tidal dephasing could be marginally detected by the next-generation detectors, as their values of $\mathcal{M}\times \rm SNR^2$ can be close to unity depending on the source parameters [\cref{fig:dfr}].


\begin{acknowledgments}
This work is supported by NSFC Grant No. 12203017 and the National SKA Program of China No. 2020SKA0120300 and and the
Fundamental Research Funds for the Central Universities (YCJJ20242227). D. G. is supported by Graduate Short-Term Study Abroad Program of Huazhong University of Science and Technology.
Z. M. is supported by the Postdoctoral Innovation Talent Support Program of CPSF (No. BX20240223) and the CPSF funded project (No. 2024M761948).
\end{acknowledgments}

\bibliography{apssamp}

\end{document}